# Magnetic proximity effect in biphenylene monolayer from first-principles

Diego López-Alcalá[a] and José J. Baldoví[a,*]

[a]*Instituto de Ciencia Molecular, Universitat de València, Catedrático José Beltrán 2, 46980 Paterna, Spain.*

On-surface chemistry has emerged as a key technique for designing novel low-dimensional materials, enabling precise manipulation of their electronic and magnetic properties at the atomic scale. It also proves highly effective for the fabrication of heterostructures. Leveraging these benefits, herein, we perform a first principles study of the magnetic proximity effect (MPE) in a heterostructure formed by a monolayer of the two-dimensional carbon allotrope biphenylene network (BPN) deposited on the surface of the above-room-temperature ferrimagnet yttrium iron garnet (YIG). Our results reveal strong hybridization between BPN orbitals and YIG surface states, resulting in non-homogeneous electron transfer and robust MPE. The proposed methodology accurately describes YIG magnetic interactions, allowing us to study the tuning effects of BPN on the magnetic properties of the substrate for the first time. Additionally, we explore the impact of van der Waals (vdW) distance at the interface, finding enhanced spin splitting up to 30% under external pressure. These findings highlight a promising strategy for inducing spin polarization in BPN without chemical modifications, opening new possibilities for BPN-based spintronic devices through the creation of heterostructures with magnetic materials.

**Introduction**

Graphene has become the flagship of two-dimensional (2D) materials since its discovery[1] because its outstanding electronic, mechanical and thermal properties.[2,3] Besides, these properties can be easily tuned by many different strategies such as the creation of heterostructures via van der Waals (vdW) stacking,[4] strain engineering,[5] electrostatic doping,[6] atom adsorption[7,8] or defect creation.[9] These exciting possibilities fueled the community to search for new 2D materials[10] and, more recently, carbon-based 2D materials such as phagraphene,[11] graphane[12] or graphullerene,[13,14] with some analogous properties to graphene, leading to a new generation of 2D π-conjugated materials.

In this context, the biphenylene network (BPN) is one of the latest 2D carbon allotrope to be synthesized.[15] This C sp$^2$ network is a metallic planar system with high stability and mechanical anisotropy,[16] and additional properties as anisotropic thermal transport,[17] lithiation[18] and hydrogenation.[19] Many theoretical studies have confirmed a prominent negative thermal expansion,[20] topological ordering and anisotropic charge transport[21,22] which makes BPN a promising new 2D C π-conjugated network for the implementation in many cutting-edge research fields.

Notwithstanding the fascinating properties of 2D carbon-based materials, many efforts have been focused on inducing magnetism in them for their application in advanced technologies based on spintronics.[23] Among them, the creation of magnetic defects,[24] the addition of magnetic dopants[25,26] or the adsorption of magnetic atoms and molecules,[27–29] have been successfully implemented. Lately, new methodologies have allowed to grow graphene on magnetic surfaces, which provides efficient spin injection on the π-conjugated C system,[30–33] thus opening an alternative route to induce magnetism in 2D C sp$^2$ materials via magnetic proximity effect (MPE). Indeed, many experimental studies have reported MPE in non-magnetic 2D materials forming vdW heterostructures with magnetic monolayers.[34,35] This effect has been extensively studied in graphene since MPE induces spin splitting on the different spin components of electronic structure, with potential applications in spin filtering or spin-dependent tunneling.[36–39] However, to the best of our knowledge, this phenomenon has not been investigated in BPN and deserves urgent attention, owing to the promising combination of the electronic properties of BPN with the emerging opportunities of magnetism at the 2D limit.[40]

In this work, we study MPE and its effects on the electronic and magnetic properties of BPN through a combination of first-principles based on Hubbard-corrected density functional theory (DFT+U), tight-binding and atomistic simulations. As a magnetic counterpart we use yttrium iron garnet ($Y_3Fe_5O_{12}$, YIG), which is a well-known insulating ferrimagnet with an above-room-temperature Curie temperature ($T_C$) of 570 K, large spin-wave lifetime and particularly low magnetic Gilbert damping.[41] Then, we rationalize, for the first time, the tunability of magnetic exchange interactions in YIG due to the proximity to another material. Furthermore, we investigate the evolution of MPE as a function of vdW distance between BPN and YIG, as it can be reduced experimentally by applying external hydrostatic pressure.[42,43] These findings shed light on the interactions at the BPN/YIG interface in view of recent developments on surface grown techniques, and pave the way to the possibility of engineering the magnetization of the single-layer BPN network.

**Computational details**

For the construction of the BPN/YIG heterostructure we combined a hexagonal 1x1x2 supercell of YIG (111) with a hexagonal supercell that contains 16 conventional unit cells of BPN. We used the *CellMatch* python code[44] that finds the most suitable combination to minimize the mismatch between both unit cells. We added a vacuum distance of 15 Å in the *z* direction to avoid interaction between both non-periodic sides of the slab. First-principles DFT+U calculations were performed using SIESTA code.[45,46] We used GGA+PBE method to describe the exchange correlation energy.[47] Hubbard U corrections (U = 7 eV) as implemented in SIESTA[48] were considered for the strongly correlated Fe 3d electrons. We used norm-conserving scalar relativistic pseudo-potentials taken from the Pseudo-Dojo database[49] in the psml format.[50] Grimme D2 dispersion corrections were applied to consider for vdW interactions.[51] A real-space mesh cutoff of 700 Ry and a 2x2x1 Monkhorst−Pack k-point mesh was used in all calculations, in combination with double-ζ basis set for all atoms. We employed diffuse functions in the surface atoms to improve the description of the interface.[52] To account for the electric field created by the asymmetric nature of the slab, we added a dipole correction as implemented in SIESTA.[53] All structures were relaxed until the forces were less than 0.04 eV/Å in all atomic coordinates. Charge transfer analysis was performed using Bader charge partition as proposed by the Henkelman group.[54] Magnetic exchange couplings (J) were computed using Green's function method as implemented in TB2J code.[55]

**Results and discussion**

A single layer of BPN is formed by the arrangement of C $sp^2$ atoms in such a way that four-, six- and eight-membered are present in the atomic thin layer. Figure 1a shows the structure of the orthorhombic conventional unit cell of BPN. Analogous to graphene, the disposition of the C atoms leads to a π-conjugated polymer, where the $p_z$ orbitals of each atom participate in the delocalized electronic π system. The band structure of the single layer BPN is shown in Figure 1b, where we can observe highly dispersive bands compatible with a high electron mobility and a metallic ground state, in agreement with previous experimental and theoretical findings.[21] Furthermore, several Dirac points can be observed in the electronic band structure. Due to its proximity to the Fermi level, the type-II Dirac cone between Γ and X at ~0.5 eV has attracted most of the attention.[56,57] BPN has been predicted to have an open-shell multiradical character in the ground state[22] that is confirmed by our calculations, in which the open-shell multiradical antiferromagnetic (AFM) configuration is 2 meV more stable than the closed-shell spin configuration (See Table S1 for further details).

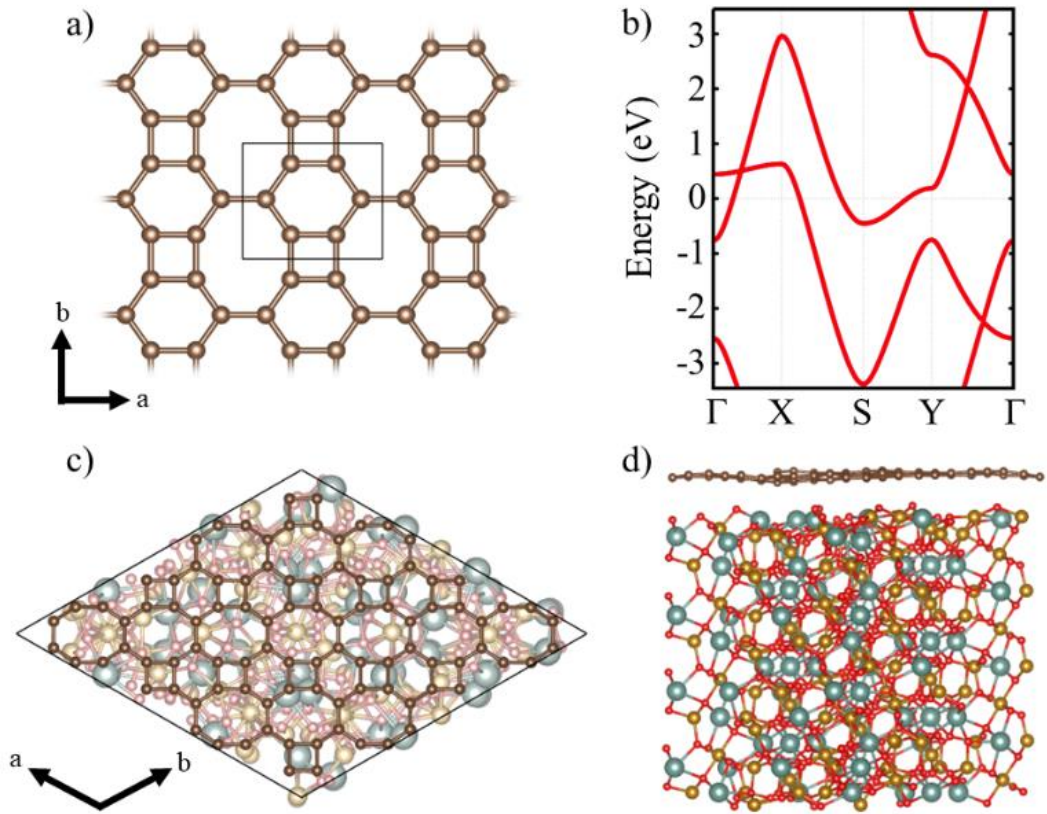

**Figure 1: a)** BPN monolayer structure and its conventional unit cell. **b)** Calculated electronic band structure of BPN monolayer. **c)** Top and **d)** lateral view of the BPN – (111) YIG surface heterostructure. Color code: yttrium (cyan), iron (golden), oxygen (red) and carbon (brown).

In order to explore the MPE caused by the interaction between an atomic thin layer of BPN and the (111) YIG surface, we create a heterostructure formed by a BPN hexagonal supercell (see Computational details) placed above a ~20 Å thick (111) YIG surface slab (Figure 1c and d). We use an oxygen termination of the slab, since previous reports on graphene/YIG (111) have confirmed that this configuration is the most stable energetically.[58] The deposition of the BPN sheet onto the magnetic surface causes many deformations in the C network, mainly due to the interaction between the C $p_z$ orbitals (pointing towards the surface) and the surface atoms of YIG with dangling bonds because of the surface reconstruction. Larger interaction can be found in the areas where the C atoms of BPN are close to the surface Y atoms of YIG. At these points, the BPN sheet shows a slightly higher tendency to approach the surface, where there is a sizeable interaction between the π electrons of BPN and the Y atoms. Our charge transfer study (Figure 2) confirms this, in which one can observe that the charge density flows towards the C atoms above the Y surface atom. This is due to a higher electronegativity of C atoms with respect to Y. At the points where the outermost atoms of the YIG surface are O we observe the opposite tendency, a depletion of the charge density in BPN. Analyzing separately the charge transfer flow at the interface it is noticeable that the charge accumulation on BPN lies down on the σ bonds at the C $sp^2$ atoms, whereas the charge depletion on the BPN mainly arises from the $p_z$ orbitals (Figure 2b and c). Due to the complexity of the YIG surface, we observe a non-homogeneous charge transfer between the BPN and the magnetic surface with an absolute number of 0.58 $e$/f.u. (f.u. = formula unit) flowing from the C network to the outermost atoms of (111) YIG surface. Our DFT-D2 calculations reveal a minimum of 2.75 Å distance between BPN and YIG surface where there is a Y-C proximity and an average of 3.22 Å, supporting the idea of a typical vdW interaction.

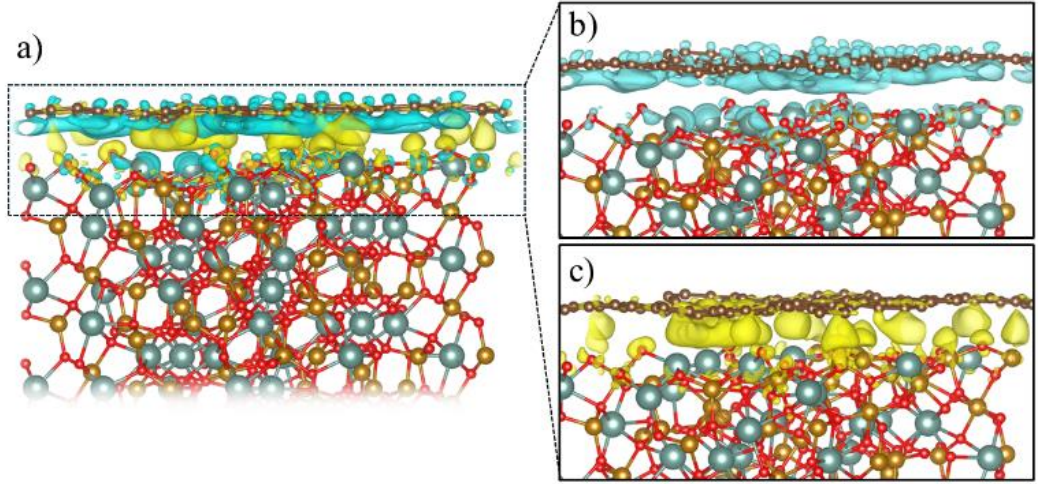

**Figure 2**: **a)** Charge density difference after BPN deposition on (111) YIG surface. Color code: blue (yellow) regions represent charge depletion (accumulation). **b)** and **c)** Show separately the charge depletion and accumulation, respectively. Isosurface is set to 0.001 $e$/Å$^3$.

Additionally, we check the stability of the heterostructure by calculating the binding energy of the system ($E_b$), $E_b = -[(E_H - E_{YIG} - E_{BP})/A]$, where A is the area of the supercell, $E_H$ is the calculated energy of the heterostructure, $E_{YIG}$ and $E_{BP}$ are the energy of the isolated (111) YIG and BPN monolayer, respectively. Our results show a binding energy of 0.3 eV/Å$^2$, which is compatible with previous theoretical studies of BPN and graphene heterostructures.[59,60] Figures 3b and c show the calculated density of states (DOS) and band structure obtained within the GGA+U approximation for the relaxed heterostructure. Firstly, we shall point out that YIG slab has a well converged thickness because the bulk-like atoms in the slab, i.e. those atoms in the inner part of the slab, can describe the well-known band gap of bulk YIG, as can be seen in the DOS of the heterostructure (Figure 3c). Although GGA is commonly underestimating the electronic band gap in many insulating systems,[61,62] bulk region of the slab shows a 2.1 eV band gap which is close to the reported experimental value (2.85 eV).[63,64] Besides, the obtained values of magnetic moments in Fe atoms belonging to the bulk-like region of the slab are 4.57 and 4.53 $\mu_B$ for octahedral and tetragonal coordinated Fe atoms, respectively. These values are in good agreement with neutron diffraction and electronic magnetic circular dichroism experiments.[65,66] Arising from the quenched orbital moment of Fe 3d orbitals, YIG possess a negligible magnetic anisotropy as reported in bibliography.[67] Above the states that represent the band gap of the bulk-like region, there are some states belonging to the surface states of YIG. This is mainly due to the dangling bonds that appear at the surface after the surface reconstruction. The states responsible for the BPN atoms are highly hybridized with the surface states of YIG near the Fermi level because of the interaction between the p$_Z$ orbitals of C atoms and the surface states of YIG, as expected because of the complex charge transfer interaction in the interface. The calculated electronic band structure with atomic contribution (Figure 3b) confirms that the energy levels crossing the Fermi level show a clear BPN-YIG hybridized character. Here, one can observe that highly dispersive bands of BPN do not show a prominent interaction with YIG electronic bands due to the described weak vdW interaction present at the interface.

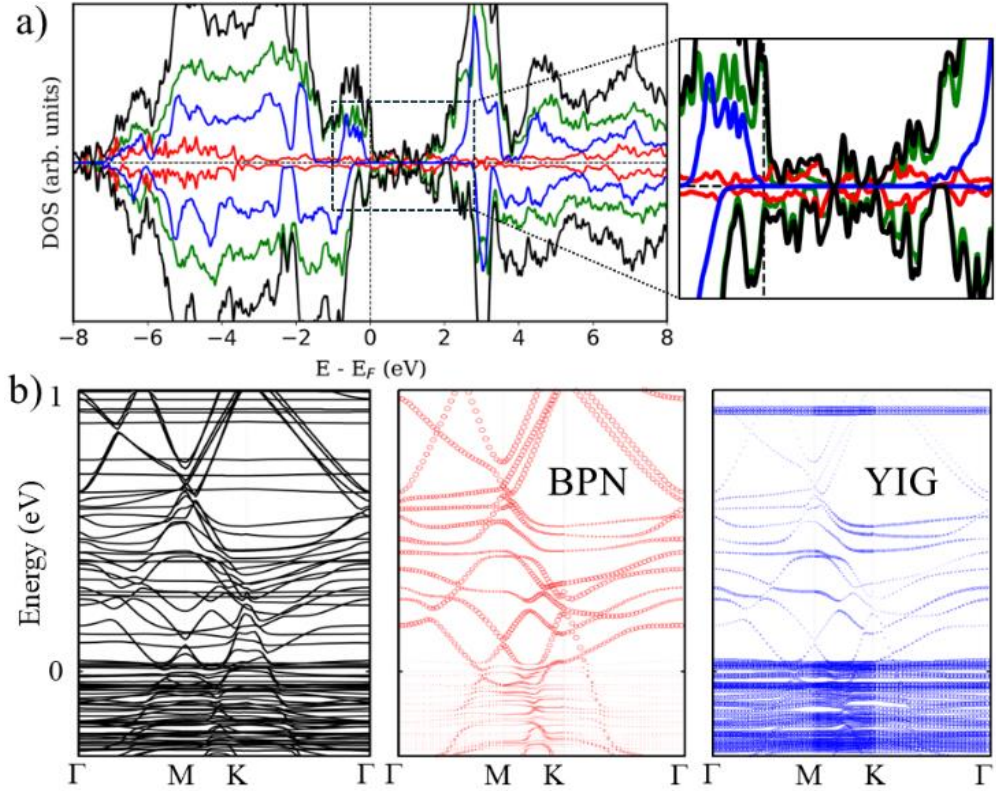

**Figure 3**: **a)** Density of states (DOS) of the BPN – YIG heterostructure. Color code: Total DOS (black), YIG bulk-like atoms (blue), YIG surface atoms (green) and BPN C atoms (red). **b)** Electronic band structure of the BPN – YIG heterostructure (black lines). The red (blue) color indicates the contribution of the BPN (YIG) atoms.

We turn now to the description of the MPE caused by the YIG surface. As BPN has been predicted to have an open-shell multiradical character at the ground state, the study of how the redistribution of spin-polarized electronic density arising from the proximity of a magnetic surface is particularly interesting in the case of BPN/YIG heterostructure. Our DFT calculations on a free standing BPN sheet reveal that C $sp^2$ atoms have an average net magnetic moment of ~0.0001 $\mu_B$. The corresponding spin density of these magnetic moments is mainly localized on the $p_Z$ orbitals of C atoms (Figure S1). After deposition of BPN on the YIG magnetic surface there is an increase in the average magnetic moment on the C atoms to 0.005 $\mu_B$/atom, which is compatible with a solid MPE interaction. This enhancement of atomic magnetic moments represents a robust insight into inducing magnetism on BPN. We calculate the spin-polarization (SP) over the energy levels of BPN as it can reveal the influence of the polarized YIG surface levels the C network. The SP can be attributed to the differences in the energy levels between the spin up and down components considered in our spin-polarized DFT calculations as: $SP = \left|\frac{N_\uparrow(E) - N_\downarrow(E)}{N_\uparrow(E) + N_\downarrow(E)}\right|$, where $N_\uparrow(E)$ and $N_\downarrow(E)$ are the density of states at a given energy for each spin component. Figure 4a shows the SP of the BPN and YIG surface states at a given range of energy (red line and green dotted line, respectively), where a modest influence of the polarization of the surface states of YIG on BPN energy levels can be observed where there are some overlaps of these energy states compatible with a weak vdW interaction. YIG surface spin-polarized states interact with the $p_Z$ orbitals of BPN at the interface leading to a hybridization of their electronic states, which causes the polarization of the BPN structure.

Magnetic Fe atoms in YIG can be differentiated into octahedrally ($Fe^O$) and tetrahedrally ($Fe^T$) coordinated (Figure 4b). Neutron-diffraction measurements have revealed that $Fe^O$ and $Fe^T$ atoms are coupled into an antiparallel configuration.[65] Since there is a $Fe^O/Fe^T$ 2:3 ratio on YIG per formula unit, a net magnetization is observed. Hence, YIG is a magnetically soft insulator that can be described using a Heisenberg model ($E = -\sum_{i \neq j} J_{ij} S_i \cdot S_j$). In our DFT calculations, we defined both spin contributions as $Fe^O$ (spin up) and $Fe^T$ (spin down). We perform a surface reconstruction of the magnetic substrate to construct the slab, which leads to a change on the disposition of the atoms exposed to the interface. This process changes the electronic and magnetic properties of the atoms close to this surface because of the presence of dangling bonds and geometrical distortions arising from the environmental changes near the surface.

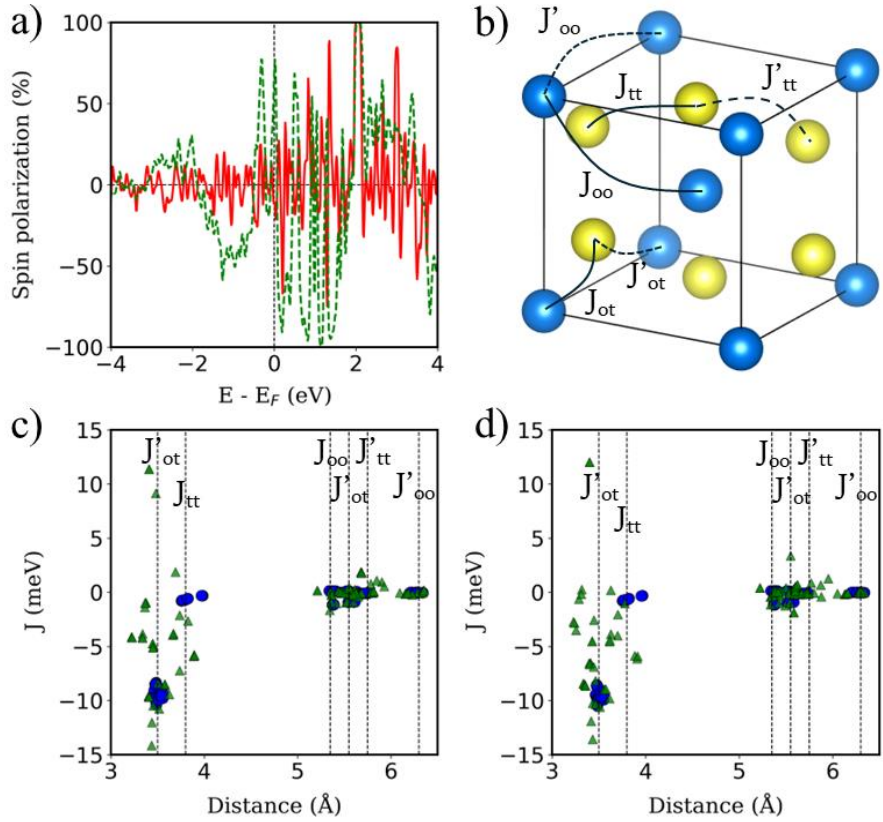

**Figure 4: a)** Polarization of the energy levels of BPN (red line) and surface YIG (dotted green line). **b)** 1/8 of YIG conventional unit cell. Only Fe atoms are shown for clarity, where blue (yellow) atoms represent the Fe octahedrally (tetrahedrally) coordinated. Solid (dashed) lines represent the nearest (next-nearest) J for each Fe. **c)** J of bulk (blue) and surface (green) atoms on the clean YIG surface. **d)** Same as c) but with BPN.

YIG is a magnetic insulator but as shown in Figure 3a, its surface has an intrinsic metallic character. Consequently, we expect the magnetic properties in the surface region of our slab to be different to those observed in the bulk-like region. To explore the magnetic properties of the heterostructure and how this interaction affects both BPN and YIG, we compute the magnetic exchange couplings (J) on the system using a computational method based on Green's functions (See Figure S2). Our results for J on the magnetic atoms present in the inner region of the YIG slab are compatible with those reported in different theoretical studies of J in bulk YIG,[68–70] validating our proposed methodology (See Supporting Information for more details). Figure 4c shows the J for magnetic Fe atoms in the bulk-like region (blue dots) and in the surface region (green triangles) of a clean YIG slab. Here we can clearly observe some similarities in the calculated J, but different new values emerge in the surface region due to the surface states arising from the surface reconstruction. The new distribution of charge density at the surface of the slab and structural rearrangements causes that Fe atoms in this region develop new interactions between them, giving rise to these magnetic interactions different than those observed in the bulk-like regions. We observe a common trend in the changes of magnetic moments when comparing the values at the surface with those at deeper Fe atoms, where the surface atoms carry a lower magnetic moment rather than those at the surface (See Table S2). These changes in the magnetic moments and the structural changes caused by surface reconstruction led to a complete change of the magnetic behavior of the surface states of YIG rather than the bulk-like states. We observe a different coordination in some Fe atoms exposed to the surface, i.e. from a tetrahedral to an octahedral coordination, hence the changes on the electronic environment near the d orbitals lead to different magnetic interactions at the surface of the slab. In this scenario, now we explore the changes in magnetic interactions that could induce the proximity of BPN given that we observe a non-negligible charge transfer at the interface. Figure 4d shows the calculated J when the BPN monolayer is placed on top of the YIG surface. Here it is clearly observed that the J of surface atoms are shifted towards lower distances due to the compression that BPN induces on the outermost part of the YIG slab. Moreover, we observe a few changes in some J, e.g., $J_{ot}'$ and $J_{tt}$, between surface atoms mainly due to the charge transfer and therefore a new charge density redistribution in this area. Regarding the changes in the calculated J in the bulk-like region we observe similar trends as before adsorption, but there is a general enhancement of the magnetic interactions in a range of 4 to 1 % (Table S3), mainly due to the charge accumulation present in YIG because of the interaction with the π-electron system of BPN.

Arising from the weak vdW interaction, the application of an external pressure may be a powerful tool to tune the distance between the BPN and the magnetic surface. This can lead to an enhancement of the MPE. Figure 5a shows a description of the charge transfer process as a function of the vdW distance. As abovementioned, at the equilibrium distance, the BPN is transferring ~0.6 *e*/f.u. to the YIG surface, which corresponds to the maximum charge transfer in the heterostructure. When we separate both components, there is a decrease of the amount of electronic density mainly due to a diminution of the overlap between the $p_Z$ orbitals of BPN and the surface states of YIG. Interestingly, we observe a change on the electronic density flow at ~d-$d_0$ = -0.8 Å, as the electrons start to flow towards the BPN. This fact has previously been reported for similar graphene heterostructures.[71] We took a deeper look at this behavior by analyzing the charge transfer difference as in Figure 2, but in this case, we compute it at the vdW distance that creates these changes (Figure S3). We find that the BPN is close enough to the surface to subtract the electronic density of the less electronegative Fe atoms exposed to the surface. This fact is crucial to understand the change in the magnitude of the charge transfer and it provides a pathway to a selective flow of the electronic density by applying external pressure.

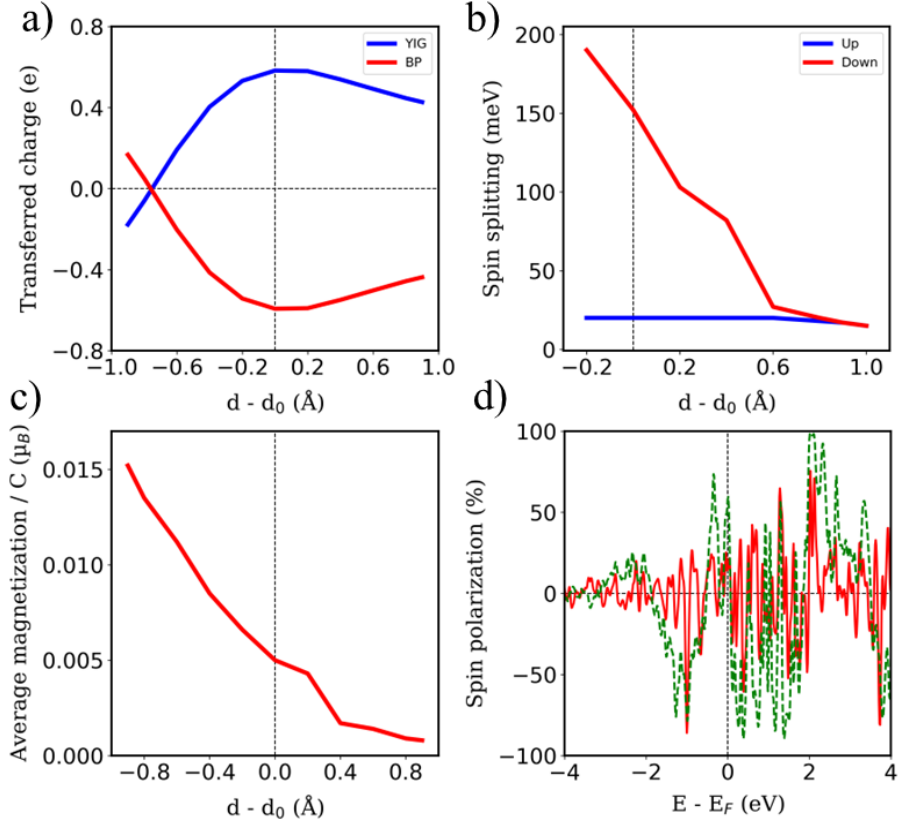

**Figure 5:** Changes on **a)** transferred charge, **b)** spin-dependent band gaps and **c)** absolute magnetization of BPN in function of the vdW distance. **(d)** Polarization of the energy levels of BPN (red line) and surface YIG (dotted green line) at $d - d_0 = -0.9$ Å.

The selected O termination of the (111) YIG surface causes that the outermost polarized atoms of the slab are $Fe^T$ (spin down), so the interaction with BPN is clearly influenced by this polarization of the interface. In this situation, a different spin splitting of the BPN bands is expected due to the polarized YIG surface. Figure 5b shows the spin-dependent band gap for up and down spin components of BPN bands at different vdW distances. A clear influence of the tetrahedral Fe atoms near the interface is clearly observed since the spin down bands are the most splitted, whereas the spin up bands are barely affected. At the equilibrium distance we observe a spin splitting of 130 meV ($\Delta_\uparrow - \Delta_\downarrow$), which is slightly higher compared to those values calculated for graphene in contact with magnetic insulators.[72,73] At large vdW distances, i.e. $d-d_0 = -1$ Å, one can see how the band gap for each spin component is equal since the MPE in this scenario is almost negligible, whereas the spin splitting of the spin components increases by 30% due to large interaction between YIG and BPN at closer vdW distances (Figure S4). Figure 5c shows the effect of the vdW distance on the BPN magnetization, where an abrupt enhancement of magnetic moments on C atoms is found at closer distances than equilibrium. This augmentation of MPE could be explained by a higher intensity of the interaction between the C atoms and the surface magnetic Fe atoms of YIG, as it can be observed in the polarization of the energy levels of BPN and surface YIG atoms at $d-d_0 = -0.9$ Å (Figure 5d) where an enhancement of the agreement into the polarization of the energy levels is noticeable in comparison to the calculated at the equilibrium distance (Figure 4a). The emergent interaction between spin-polarized Fe atoms and BPN at lower vdW distances causes that the polarization of C atoms starts to be dependent on the region of the surface as it has complex morphology, and outermost magnetic atoms are not symmetrically distributed across the surface. As YIG surface atoms have different SP the enhancement of the interaction between these magnetic atoms and BPN causes different SP patterns on BPN. Figure 6 shows the difference between the spin density of BPN at the equilibrium distance and $d - d_0 = -0.9$ Å, where a change in the magnetic domains is noticeable clearly influenced by the YIG surface magnetic states. These findings strongly demonstrate that

YIG can induce MPE on BPN, which lays the groundwork for the potential induction of π magnetism within the carbon sp$^2$ network.

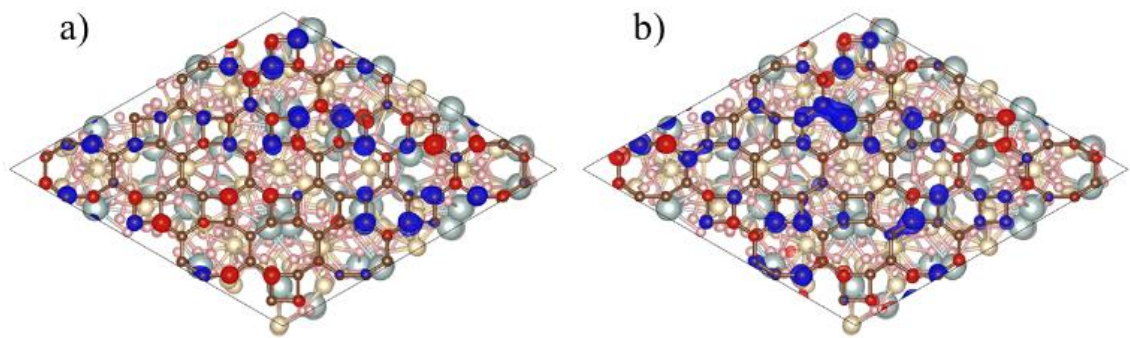

**Figure 6:** Spin density of BPN at **a)** equilibrium distance and **b)** at d – d$_0$ = – 0.9 Å. Isosurface are set to 0.0005 and 0.0015 $e$/Å$^3$, respectively. Color code: spin density up (blue) and down (red).

**Conclusions**

In this work, we present a detailed theoretical study of MPE on the recently synthesized graphene allotrope biphenylene network deposited on a magnetic surface for the first time. Our results reveal robust MPE on the 2D C sp$^2$ network, with an enhancement of magnetization in the monolayer up to ~0.02 $\mu_B$/C atom, due to the intrinsic ferrimagnetic character of YIG (111). On the other hand, the inhomogeneity of the YIG (111) surface provides different charge transfer in the interface, which creates a local tuning of the magnetic interactions in YIG. The presence of the π electron cloud present in the C sp$^2$ atoms induces a general enhancement in the range 1–4% of the magnetic exchange interactions in the bulk-like region of YIG. Finally, we investigated the effect of the vdW distance on the electronic and magnetic properties of the system, showing that the induced magnetism on BPN can be tuned by 200%. Besides, we found a 130 meV spin splitting of the bands that could be enhanced up to 30% by applying an external pressure to the heterostructure. These findings hold potential for applications in spin-selective transport and its implementation in cutting-edge spintronic-based devices, as BPN exhibits several interesting properties that can be enhanced by MPE.

**Acknowledgements**

The authors acknowledge the financial support from the European Union (ERC-2021-StG101042680 2D-SMARTiES) amd the Generalitat Valenciana (grant CIDEXG/2023/1).

**Notes and references**

(1) Novoselov, K. S.; Geim, A. K.; Morozov, S. V.; Jiang, D.; Zhang, Y.; Dubonos, S. V.; Grigorieva, I. V.; Firsov, A. A. Electric Field Effect in Atomically Thin Carbon Films. *Science (1979)* **2004**, *306* (5696), 666–669.
(2) Geim, A. K.; Novoselov, K. S. The Rise of Graphene. *Nat Mater* **2007**, *6* (3), 183–191.
(3) Castro Neto, A. H.; Guinea, F.; Peres, N. M. R.; Novoselov, K. S.; Geim, A. K. The Electronic Properties of Graphene. *Rev Mod Phys* **2009**, *81* (1), 109–162.
(4) Solís-Fernández, P.; Bissett, M.; Ago, H. Synthesis, Structure and Applications of Graphene-Based 2D Heterostructures. *Chem Soc Rev* **2017**, *46* (15), 4572–4613.
(5) Choi, S.-M.; Jhi, S.-H.; Son, Y.-W. Effects of Strain on Electronic Properties of Graphene. *Phys Rev B* **2010**, *81* (8), 081407.
(6) Giovannetti, G.; Khomyakov, P. A.; Brocks, G.; Karpan, V. M.; van den Brink, J.; Kelly, P. J. Doping Graphene with Metal Contacts. *Phys Rev Lett* **2008**, *101* (2), 026803.
(7) Lazar, P.; Karlický, F.; Jurečka, P.; Kocman, M.; Otyepková, E.; Šafářová, K.; Otyepka, M. Adsorption of Small Organic Molecules on Graphene. *J Am Chem Soc* **2013**, *135* (16), 6372–6377.

(8) Kong, L.; Enders, A.; Rahman, T. S.; Dowben, P. A. Molecular Adsorption on Graphene. *Journal of Physics: Condensed Matter* **2014**, *26* (44), 443001.
(9) Banhart, F.; Kotakoski, J.; Krasheninnikov, A. V. Structural Defects in Graphene. *ACS Nano* **2011**, *5* (1), 26–41.
(10) Novoselov, K. S.; Mishchenko, A.; Carvalho, A.; Castro Neto, A. H. 2D Materials and van Der Waals Heterostructures. *Science (1979)* **2016**, *353* (6298).
(11) Wang, Z.; Zhou, X.-F.; Zhang, X.; Zhu, Q.; Dong, H.; Zhao, M.; Oganov, A. R. Phagraphene: A Low-Energy Graphene Allotrope Composed of 5–6–7 Carbon Rings with Distorted Dirac Cones. *Nano Lett* **2015**, *15* (9), 6182–6186.
(12) Zhou, C.; Chen, S.; Lou, J.; Wang, J.; Yang, Q.; Liu, C.; Huang, D.; Zhu, T. Graphene's Cousin: The Present and Future of Graphane. *Nanoscale Res Lett* **2014**, *9* (1), 26.
(13) Hou, L.; Cui, X.; Guan, B.; Wang, S.; Li, R.; Liu, Y.; Zhu, D.; Zheng, J. Synthesis of a Monolayer Fullerene Network. *Nature* **2022**, *606* (7914), 507–510.
(14) Meirzadeh, E.; Evans, A. M.; Rezaee, M.; Milich, M.; Dionne, C. J.; Darlington, T. P.; Bao, S. T.; Bartholomew, A. K.; Handa, T.; Rizzo, D. J.; Wiscons, R. A.; Reza, M.; Zangiabadi, A.; Fardian-Melamed, N.; Crowther, A. C.; Schuck, P. J.; Basov, D. N.; Zhu, X.; Giri, A.; Hopkins, P. E.; Kim, P.; Steigerwald, M. L.; Yang, J.; Nuckolls, C.; Roy, X. A Few-Layer Covalent Network of Fullerenes. *Nature* **2023**, *613* (7942), 71–76.
(15) Fan, Q.; Yan, L.; Tripp, M. W.; Krejčí, O.; Dimosthenous, S.; Kachel, S. R.; Chen, M.; Foster, A. S.; Koert, U.; Liljeroth, P.; Gottfried, J. M. Biphenylene Network: A Nonbenzenoid Carbon Allotrope. *Science (1979)* **2021**, *372* (6544), 852–856.
(16) Luo, Y.; Ren, C.; Xu, Y.; Yu, J.; Wang, S.; Sun, M. A First Principles Investigation on the Structural, Mechanical, Electronic, and Catalytic Properties of Biphenylene. *Sci Rep* **2021**, *11* (1), 19008.
(17) Veeravenkata, H. P.; Jain, A. Density Functional Theory Driven Phononic Thermal Conductivity Prediction of Biphenylene: A Comparison with Graphene. *Carbon N Y* **2021**, *183*, 893–898.
(18) Lherbier, A.; Vander Marcken, G.; Van Troeye, B.; Botello-Méndez, A. R.; Adjizian, J.-J.; Hautier, G.; Gonze, X.; Rignanese, G.-M.; Charlier, J.-C. Lithiation Properties of $sp^2$ Carbon Allotropes. *Phys Rev Mater* **2018**, *2* (8), 085408.
(19) Liao, Y.; Shi, X.; Ouyang, T.; Li, J.; Zhang, C.; Tang, C.; He, C.; Zhong, J. New Two-Dimensional Wide Band Gap Hydrocarbon Insulator by Hydrogenation of a Biphenylene Sheet. *J Phys Chem Lett* **2021**, *12* (36), 8889–8896.
(20) Li, Q.; Zhou, J.; Liu, G.; Wan, X. G. Extraordinary Negative Thermal Expansion of Monolayer Biphenylene. *Carbon N Y* **2022**, *187*, 349–353.
(21) Son, Y.-W.; Jin, H.; Kim, S. Magnetic Ordering, Anomalous Lifshitz Transition, and Topological Grain Boundaries in Two-Dimensional Biphenylene Network. *Nano Lett* **2022**, *22* (7), 3112–3117.
(22) Alcón, I.; Calogero, G.; Papior, N.; Antidormi, A.; Song, K.; Cummings, A. W.; Brandbyge, M.; Roche, S. Unveiling the Multiradical Character of the Biphenylene Network and Its Anisotropic Charge Transport. *J Am Chem Soc* **2022**, *144* (18), 8278–8285.
(23) Han, W.; Kawakami, R. K.; Gmitra, M.; Fabian, J. Graphene Spintronics. *Nat Nanotechnol* **2014**, *9* (10), 794–807.
(24) Yazyev, O. V.; Helm, L. Defect-Induced Magnetism in Graphene. *Phys Rev B* **2007**, *75* (12), 125408.
(25) Zhao, X.; Wang, T.; Xia, C.; Dai, X.; Wei, S.; Yang, L. Magnetic Doping in Two-Dimensional Transition-Metal Dichalcogenide Zirconium Diselenide. *J Alloys Compd* **2017**, *698*, 611–616.
(26) Kochat, V.; Apte, A.; Hachtel, J. A.; Kumazoe, H.; Krishnamoorthy, A.; Susarla, S.; Idrobo, J. C.; Shimojo, F.; Vashishta, P.; Kalia, R.; Nakano, A.; Tiwary, C. S.; Ajayan, P. M. Re Doping in 2D Transition Metal Dichalcogenides as a New Route to Tailor Structural Phases and Induced Magnetism. *Advanced Materials* **2017**, *29* (43).
(27) González-Herrero, H.; Gómez-Rodríguez, J. M.; Mallet, P.; Moaied, M.; Palacios, J. J.; Salgado, C.; Ugeda, M. M.; Veuillen, J.-Y.; Yndurain, F.; Brihuega, I. Atomic-Scale Control of Graphene Magnetism by Using Hydrogen Atoms. *Science (1979)* **2016**, *352* (6284), 437–441.

(28) Qiao, Z.; Yang, S. A.; Feng, W.; Tse, W.-K.; Ding, J.; Yao, Y.; Wang, J.; Niu, Q. Quantum Anomalous Hall Effect in Graphene from Rashba and Exchange Effects. *Phys Rev B* **2010**, *82* (16), 161414.
(29) Ding, J.; Qiao, Z.; Feng, W.; Yao, Y.; Niu, Q. Engineering Quantum Anomalous/Valley Hall States in Graphene via Metal-Atom Adsorption: An *Ab-Initio* Study. *Phys Rev B* **2011**, *84* (19), 195444.
(30) Wang, Z.; Tang, C.; Sachs, R.; Barlas, Y.; Shi, J. Proximity-Induced Ferromagnetism in Graphene Revealed by the Anomalous Hall Effect. *Phys Rev Lett* **2015**, *114* (1), 016603.
(31) Leutenantsmeyer, J. C.; Kaverzin, A. A.; Wojtaszek, M.; van Wees, B. J. Proximity Induced Room Temperature Ferromagnetism in Graphene Probed with Spin Currents. *2d Mater* **2016**, *4* (1), 014001.
(32) Evelt, M.; Ochoa, H.; Dzyapko, O.; Demidov, V. E.; Yurgens, A.; Sun, J.; Tserkovnyak, Y.; Bessonov, V.; Rinkevich, A. B.; Demokritov, S. O. Chiral Charge Pumping in Graphene Deposited on a Magnetic Insulator. *Phys Rev B* **2017**, *95* (2), 024408.
(33) Singh, S.; Katoch, J.; Zhu, T.; Meng, K.-Y.; Liu, T.; Brangham, J. T.; Yang, F.; Flatté, M. E.; Kawakami, R. K. Strong Modulation of Spin Currents in Bilayer Graphene by Static and Fluctuating Proximity Exchange Fields. *Phys Rev Lett* **2017**, *118* (18), 187201.
(34) Zollner, K.; Faria Junior, P. E.; Fabian, J. Proximity Exchange Effects in $MoSe_2$ and $WSe_2$ Heterostructures with $CrI_3$: Twist Angle, Layer, and Gate Dependence. *Phys Rev B* **2019**, *100* (8), 085128.
(35) Tang, C.; Zhang, Z.; Lai, S.; Tan, Q.; Gao, W. Magnetic Proximity Effect in Graphene/$CrBr_3$ van Der Waals Heterostructures. *Advanced Materials* **2020**, *32* (16).
(36) Zollner, K.; Gmitra, M.; Frank, T.; Fabian, J. Theory of Proximity-Induced Exchange Coupling in Graphene on HBN/(Co, Ni). *Phys Rev B* **2016**, *94* (15), 155441.
(37) Mendes, J. B. S.; Alves Santos, O.; Meireles, L. M.; Lacerda, R. G.; Vilela-Leão, L. H.; Machado, F. L. A.; Rodríguez-Suárez, R. L.; Azevedo, A.; Rezende, S. M. Spin-Current to Charge-Current Conversion and Magnetoresistance in a Hybrid Structure of Graphene and Yttrium Iron Garnet. *Phys Rev Lett* **2015**, *115* (22), 226601.
(38) Leutenantsmeyer, J. C.; Kaverzin, A. A.; Wojtaszek, M.; van Wees, B. J. Proximity Induced Room Temperature Ferromagnetism in Graphene Probed with Spin Currents. *2d Mater* **2016**, *4* (1), 014001.
(39) Singh, S.; Katoch, J.; Zhu, T.; Meng, K.-Y.; Liu, T.; Brangham, J. T.; Yang, F.; Flatté, M. E.; Kawakami, R. K. Strong Modulation of Spin Currents in Bilayer Graphene by Static and Fluctuating Proximity Exchange Fields. *Phys Rev Lett* **2017**, *118* (18), 187201.
(40) Gibertini, M.; Koperski, M.; Morpurgo, A. F.; Novoselov, K. S. Magnetic 2D Materials and Heterostructures. *Nat Nanotechnol* **2019**, *14* (5), 408–419.
(41) Cherepanov, V.; Kolokolov, I.; L'vov, V. The Saga of YIG: Spectra, Thermodynamics, Interaction and Relaxation of Magnons in a Complex Magnet. *Phys Rep* **1993**, *229* (3), 81–144.
(42) Fülöp, B.; Márffy, A.; Zihlmann, S.; Gmitra, M.; Tóvári, E.; Szentpéteri, B.; Kedves, M.; Watanabe, K.; Taniguchi, T.; Fabian, J.; Schönenberger, C.; Makk, P.; Csonka, S. Boosting Proximity Spin–Orbit Coupling in Graphene/$WSe_2$ Heterostructures via Hydrostatic Pressure. *NPJ 2D Mater Appl* **2021**, *5* (1), 82.
(43) Li, C.; Cheng, W.; Zhang, X.; Zhang, P.; Zheng, Q.; Yan, Z.; Han, J.; Dai, G.; Wang, S.; Quan, Z.; Liu, Y.; Zhu, J. Tuning of Interlayer Interaction in $MoS_2$–$WS_2$ van Der Waals Heterostructures Using Hydrostatic Pressure. *The Journal of Physical Chemistry C* **2023**, *127* (16), 7784–7791.
(44) Lazić, P. CellMatch: Combining Two Unit Cells into a Common Supercell with Minimal Strain. *Comput Phys Commun* **2015**, *197*, 324–334.
(45) Soler, J. M.; Artacho, E.; Gale, J. D.; García, A.; Junquera, J.; Ordejón, P.; Sánchez-Portal, D. The SIESTA Method for *Ab Initio* Order-$N$ Materials Simulation. *Journal of Physics: Condensed Matter* **2002**, *14* (11), 2745–2779.
(46) García, A.; Papior, N.; Akhtar, A.; Artacho, E.; Blum, V.; Bosoni, E.; Brandimarte, P.; Brandbyge, M.; Cerdá, J. I.; Corsetti, F.; Cuadrado, R.; Dikan, V.; Ferrer, J.; Gale, J.; García-Fernández, P.; García-Suárez, V. M.; García, S.; Huhs, G.; Illera, S.; Korytár, R.; Koval, P.; Lebedeva, I.; Lin, L.; López-Tarifa, P.; Mayo, S. G.; Mohr, S.; Ordejón, P.; Postnikov, A.;

Pouillon, Y.; Pruneda, M.; Robles, R.; Sánchez-Portal, D.; Soler, J. M.; Ullah, R.; Yu, V. W.; Junquera, J. Siesta: Recent Developments and Applications. *J Chem Phys* **2020**, *152* (20).
(47) Perdew, J. P.; Burke, K.; Ernzerhof, M. Generalized Gradient Approximation Made Simple. *Phys Rev Lett* **1996**, *77* (18), 3865–3868.
(48) Dudarev, S. L.; Botton, G. A.; Savrasov, S. Y.; Humphreys, C. J.; Sutton, A. P. Electron-Energy-Loss Spectra and the Structural Stability of Nickel Oxide: An LSDA+U Study. *Phys Rev B* **1998**, *57* (3), 1505–1509.
(49) van Setten, M. J.; Giantomassi, M.; Bousquet, E.; Verstraete, M. J.; Hamann, D. R.; Gonze, X.; Rignanese, G.-M. The PseudoDojo: Training and Grading a 85 Element Optimized Norm-Conserving Pseudopotential Table. *Comput Phys Commun* **2018**, *226*, 39–54.
(50) García, A.; Verstraete, M. J.; Pouillon, Y.; Junquera, J. The Psml Format and Library for Norm-Conserving Pseudopotential Data Curation and Interoperability. *Comput Phys Commun* **2018**, *227*, 51–71.
(51) Grimme, S. Semiempirical GGA-Type Density Functional Constructed with a Long-Range Dispersion Correction. *J Comput Chem* **2006**, *27* (15), 1787–1799.
(52) García-Gil, S.; García, A.; Lorente, N.; Ordejón, P. Optimal Strictly Localized Basis Sets for Noble Metal Surfaces. *Phys Rev B* **2009**, *79* (7), 075441.
(53) Bengtsson, L. Dipole Correction for Surface Supercell Calculations. *Phys Rev B* **1999**, *59* (19), 12301–12304.
(54) Henkelman, G.; Arnaldsson, A.; Jónsson, H. A Fast and Robust Algorithm for Bader Decomposition of Charge Density. *Comput Mater Sci* **2006**, *36* (3), 354–360.
(55) He, X.; Helbig, N.; Verstraete, M. J.; Bousquet, E. TB2J: A Python Package for Computing Magnetic Interaction Parameters. *Comput Phys Commun* **2021**, *264*, 107938.
(56) Liu, P.-F.; Li, J.; Zhang, C.; Tu, X.-H.; Zhang, J.; Zhang, P.; Wang, B.-T.; Singh, D. J. Type-II Dirac Cones and Electron-Phonon Interaction in Monolayer Biphenylene from First-Principles Calculations. *Phys Rev B* **2021**, *104* (23), 235422.
(57) Lage, L. L.; Arroyo-Gascón, O.; Chico, L.; Latgé, A. Robustness of Type-II Dirac Cones in Biphenylene: From Nanoribbons to Symmetric Bilayer Stacking. *Phys Rev B* **2024**, *110* (16), 165423.
(58) Sakai, S.; Erohin, S. V.; Popov, Z. I.; Haku, S.; Watanabe, T.; Yamada, Y.; Entani, S.; Li, S.; Avramov, P. V.; Naramoto, H.; Ando, K.; Sorokin, P. B.; Yamauchi, Y. Dirac Cone Spin Polarization of Graphene by Magnetic Insulator Proximity Effect Probed with Outermost Surface Spin Spectroscopy. *Adv Funct Mater* **2018**, *28* (20).
(59) Nazir, M. A.; Shen, Y.; Zhang, C.; Wang, Q. Schottky-Barrier-Free VdW Contact in a 2D Penta-NiN$_2$/Biphenylene Network Heterostructure. *The Journal of Physical Chemistry C* **2023**, *127* (50), 24452–24457.
(60) Nazir, M. A.; Shen, Y.; Hassan, A.; Wang, Q. The Electronic and Interfacial Properties of a VdW Heterostructure Composed of Penta-PdSe$_2$ and Biphenylene Monolayers. *Mater Adv* **2023**, *4* (6), 1566–1571.
(61) Sham, L. J.; Schlüter, M. Density-Functional Theory of the Energy Gap. *Phys Rev Lett* **1983**, *51* (20), 1888–1891.
(62) Perdew, J. P.; Levy, M. Physical Content of the Exact Kohn-Sham Orbital Energies: Band Gaps and Derivative Discontinuities. *Phys Rev Lett* **1983**, *51* (20), 1884–1887.
(63) Metselaar, R.; Larsen, P. K. High-Temperature Electrical Properties of Yttrium Iron Garnet under Varying Oxygen Pressures. *Solid State Commun* **1974**, *15* (2), 291–294.
(64) Wittekoek, S.; Popma, T. J. A.; Robertson, J. M.; Bongers, P. F. Magneto-Optic Spectra and the Dielectric Tensor Elements of Bismuth-Substituted Iron Garnets at Photon Energies between 2.2-5.2 EV. *Phys Rev B* **1975**, *12* (7), 2777–2788.
(65) Bouguerra, A.; Fillion, G.; Hlil, E. K.; Wolfers, P. Y3Fe5O12 Yttrium Iron Garnet and Lost Magnetic Moment (Computing of Spin Density). *J Alloys Compd* **2007**, *442* (1–2), 231–234.
(66) Song, D.; Li, G.; Cai, J.; Zhu, J. A General Way for Quantitative Magnetic Measurement by Transmitted Electrons. *Sci Rep* **2016**, *6* (1), 18489.
(67) Lin, Y.; Jin, L.; Zhang, D.; Zhang, H.; Wang, Z. Magnetic Anisotropy of Yttrium Iron Garnet from Density Functional Theory. *The Journal of Physical Chemistry C* **2023**, *127* (1), 689–695.

(68) Gorbatov, O. I.; Johansson, G.; Jakobsson, A.; Mankovsky, S.; Ebert, H.; Di Marco, I.; Minár, J.; Etz, C. Magnetic Exchange Interactions in Yttrium Iron Garnet: A Fully Relativistic First-Principles Investigation. *Phys Rev B* **2021**, *104* (17), 174401.

(69) Xie, L.-S.; Jin, G.-X.; He, L.; Bauer, G. E. W.; Barker, J.; Xia, K. First-Principles Study of Exchange Interactions of Yttrium Iron Garnet. *Phys Rev B* **2017**, *95* (1), 014423.

(70) Princep, A. J.; Ewings, R. A.; Ward, S.; Tóth, S.; Dubs, C.; Prabhakaran, D.; Boothroyd, A. T. The Full Magnon Spectrum of Yttrium Iron Garnet. *NPJ Quantum Mater* **2017**, *2* (1), 63.

(71) Cui, Q.; Liang, J.; Yang, B.; Wang, Z.; Li, P.; Cui, P.; Yang, H. Giant Enhancement of Perpendicular Magnetic Anisotropy and Induced Quantum Anomalous Hall Effect in Graphene/$NiI_2$ Heterostructures via Tuning the van Der Waals Interlayer Distance. *Phys Rev B* **2020**, *101* (21), 214439.

(72) Yang, H. X.; Hallal, A.; Terrade, D.; Waintal, X.; Roche, S.; Chshiev, M. Proximity Effects Induced in Graphene by Magnetic Insulators: First-Principles Calculations on Spin Filtering and Exchange-Splitting Gaps. *Phys Rev Lett* **2013**, *110* (4), 046603.

(73) Hallal, A.; Ibrahim, F.; Yang, H.; Roche, S.; Chshiev, M. Tailoring Magnetic Insulator Proximity Effects in Graphene: First-Principles Calculations. *2d Mater* **2017**, *4* (2), 025074.